\documentclass[a4,12pt]{article}

\usepackage[dvipdfm]{graphicx}

\title
{
An exact solution to determination of an open orbit 
}
\author
  {H.~Asada\thanks{Email: asada@phys.hirosaki-u.ac.jp}\\ 
Faculty of Science and Technology, Hirosaki University, \\
Hirosaki 036-8561, Japan}

\begin{document}

\date{\today}



\maketitle


\begin{abstract}
We present an exact solution of 
the equations for orbit determination of a two body system 
in a hyperbolic or parabolic motion. 
In solving this problem, we extend the method employed by 
Asada, Akasaka and Kasai (AAK) for a binary system 
in an elliptic orbit. 
The solutions applicable to each of elliptic, hyperbolic and
parabolic orbits are obtained by the new approach, and they 
are all expressed in an explicit form, remarkably, 
only in terms of elementary functions. 
We show also that the solutions for an open orbit 
are recovered by making a suitable transformation of the AAK solution 
for an elliptic case. 
\end{abstract}

keywords: astrometry ---  celestial mechanics ---  
orbit determination 

\section{Introduction}
Two body problems are very classical in celestial mechanics and 
have been studied thoroughly since Kepler discovered 
the laws of motion of celestial objects 
(e.g., Aitken 1964, Goldstein 1980, Danby 1988, Roy 1988, 
Murray and Dermott 1999, Beutler 2004). 
The regular orbits of a system of two masses in Newtonian mechanics 
are of three types: ellipse, parabola and hyperbola. 
The latter two cases, in which the separation of two bodies will 
become infinite in the remote future, may be called open orbits. 
The singular orbit is a linear one corresponding to a head-on collision, 
which is extremely special. 
In this paper, we consider regular orbits mentioned above: 
for a binary system, the orbit determination may bring us 
some informations about its formation and evolution mechanism. 
In an open orbit, one may 
infer an impact parameter and an initial relative velocity 
of two masses, one of which may be ejected by some explosive 
mechanism such as a supernova or by three body scattering. 
For instance, some observations reveal that kicked pulsars move 
at unusually high speed (Anderson et al. 1975, Hobbs et al. 2005). 

The orbit determination of {\it visual double stars} was solved 
first by Savary in 1827, secondly by Encke 1832, thirdly by 
Herschel 1833 and by many authors including Kowalsky, Thiele and Innes 
(Aitken 1964 for a review on earlier works; for the
state-of-the-art techniques, e.g, Eichhorn and Xu 1990, 
Catovic and Olevic 1992, Olevic and Cvetkovic 2004). 
Here, a visual binary is a system of two stars both of which can be seen. 
The relative vector from the primary star to the secondary is 
in an elliptic motion with a focus at the primary. 
This relative vector is observable because the two stars are seen. 
On the other hand, an {\it astrometric binary} is a system of 
two objects where one object can be seen but the other cannot 
like a black hole or a very dim star. In this case, it is impossible 
to directly measure the relative vector connecting the two objects, 
because one end of the separation of the binary, namely 
the secondary, cannot be seen. The measures are made in the position 
of the primary with respect to unrelated reference objects 
(e.g., a quasar) whose proper motion is either negligible or known. 

As a method to determine the orbital elements of a binary, 
an analytic solution in an explicit form has been found 
by Asada, Akasaka and Kasai (2004, henceforth AAK). 
This solution is given in a closed form by requiring 
neither iterative nor numerical methods. 
One may naturally seek an analytic method of orbit determination 
for open orbits. 
An extension for open orbits done earlier by Dommanget (1978) 
used the Thiele-Innes method, namely solved numerically 
the Kepler equation. 
As a result, the method by Dommanget does not provide 
an explicit solution in a closed form. 
Therefore, let us extend the explicit solution 
by AAK to open orbits. 
Then, we would face the following problem. 
AAK formalism uses a fact that the semimajor and semiminor axes of 
an ellipse divide it into quarters. 
This fact plays a crucial role in determining the position of 
the common center of mass on a celestial sphere; 
we should note here that the projected  common center of mass is 
not necessarily a focus of an apparent ellipse. 
The division into quarters is possible for neither
a parabola nor a hyperbola. 

The purpose of this paper is to generalize AAK approach 
so that we can treat an open orbit. 
This paper is organized as follows. 
Sec. 2 presents a generalized AAK formalism. 
In Sec. 3, the generalized approach is employed to obtain 
the method of orbit determination for a hyperbolic orbit. 
In Sec. 4, the formula for a parabolic orbit is presented. 
In Sec. 5, we recover these formulae by making a suitable 
transformation of that for an elliptic orbit 
with some limiting procedures. 
Sec. 6 is devoted to Conclusion.

\section{Generalizing AAK formalism for an elliptic motion}
\subsection{An apparent ellipse}
We denote by $(\bar x, \bar y)$ the Cartesian coordinates 
on a celestial sphere that is perpendicular to the line of sight.
A general form of an ellipse on a celestial sphere is 
\begin{equation}
\alpha \bar x^2 + \beta \bar y^2 + 2\gamma \bar x \bar y 
+ 2\delta \bar x + 2\varepsilon \bar y = 1 , 
\label{ellipse}
\end{equation}
which is characterized by five parameters; the position of its center, 
the length of its semimajor/semiminor axes and the rotational degree 
of freedom. 
By at least five measurements of the location of a star, 
one can determine all the parameters. 
Henceforth, we adopt the Cartesian coordinates $(x, y)$
such that the apparent ellipse can be reexpressed 
in the standard form as 
\begin{equation}
\frac{x^2}{a^2}+\frac{y^2}{b^2}=1 , 
\label{ellipse3}
\end{equation}
where we assume $a\geq b$ without loss of generality. 
The ellipticity, $e$, is $\sqrt{1-b^2/a^2}$. 

\subsection{The common center of mass}
A focus of the original Keplerian ellipse is not always that of 
the apparent one because of the inclination of the orbital plane. 
We should note that a focus of the original Keplerian orbit 
is the common center of mass of a binary, around which 
a component star moves at the constant-areal velocity 
following the Keplerian second law (the conservation law of 
the angular momentum in the classical mechanics). 
This enables us to find out the location of the common center of mass 
as shown below. 

A star is located at ${\bf P}_j=(x_j, y_j)=(a \cos u_j, b \sin u_j)$ 
on a celestial sphere at each epoch $t_j$ for $j=1, \cdots, 4$, 
where $t_j>t_k$ for $j>k$. 
Here, $u_j$ denotes the eccentric angle in the apparent ellipse 
but not the eccentric anomaly in the true one; 
the eccentric anomaly of the original Keplerian orbit 
is not observable. 
We assume anti-clockwise motion, such that $u_j > u_k$ 
for $j>k$.
All we must do in the case of the clockwise motion is 
to change the signature of the area in Eq. ($\ref{areaS}$) 
in the following. 
We define the time interval as $T(j, k)=t_j-t_k$. 

The common center of mass of the binary is projected 
onto the celestial sphere at 
${\bf P}_{\mbox{C}}=(x_{\mbox{c}}, y_{\mbox{c}})$. 
Even after the projection, the law of constant-areal 
velocity still holds, where we should note that the area 
is swept by the line interval between the projected common center 
and the star. 
The area swept during the time interval, $T(j, k)$, 
is denoted by $S(j, k)$. 
The total area of the observed ellipse is denoted by $S=\pi ab$. 
The law of the constant areal velocity on the celestial sphere becomes 
\begin{equation}
\frac{S}{T}=\frac{S(j, k)}{T(j, k)} ,
\label{areavelocity}
\end{equation}
where 
\begin{equation}
S(j, k)=\frac12 ab  
\Bigl[
u_j-u_k
-\frac{x_{\mbox{c}}}{a}(\sin u_j-\sin u_k)
+\frac{y_{\mbox{c}}}{b}(\cos u_j-\cos u_k) 
\Bigr] . 
\label{areaS}
\end{equation}

Equation $(\ref{areavelocity})$ is rewritten explicitly as 
\begin{eqnarray}
\frac{S(2, 1)}{T(2, 1)}&=&\frac{S(3, 2)}{T(3, 2)} , \\
\frac{S(3, 2)}{T(3, 2)}&=&\frac{S(4, 3)}{T(4, 3)} . 
\label{areavelocity1234}
\end{eqnarray}
They are solved for $x_{\mbox{c}}$ and $y_{\mbox{c}}$ as 
\begin{eqnarray}
x_{\mbox{c}}&=&-a \frac{B_1 C_2-B_2 C_1}{A_1 B_2-A_2 B_1} , 
\label{xC}\\
y_{\mbox{c}}&=&b \frac{C_1 A_2-C_2 A_1}{A_1 B_2-A_2 B_1} , 
\label{yC}
\end{eqnarray}
where 
\begin{eqnarray}
A_j&=&T(j+1, j)\sin u_{j+2}+T(j+2, j+1)\sin u_j
\nonumber\\
&&
-T(j+2, j)\sin u_{j+1} , 
\label{Aj} \\
B_j&=&T(j+1, j)\cos u_{j+2}+T(j+2, j+1)\cos u_j
\nonumber\\
&&
-T(j+2, j)\cos u_{j+1} , 
\label{Bj} \\
C_j&=&T(j+1, j) u_{j+2}+T(j+2, j+1) u_j-T(j+2, j) u_{j+1} .
\label{Cj} 
\end{eqnarray}

The periastron is projected onto the observed ellipse 
at ${\bf P}_{\mbox{A}}\equiv(x_{\mbox{A}}, y_{\mbox{A}})
=(a \cos u_{\mbox{A}}, b \sin u_{\mbox{A}})$. 
The ratio of the semimajor axis to the distance between the center and 
the focus of the ellipse remains unchanged, even after the projection. 
Hence, we find 
\begin{equation}
{\bf P}_{\mbox{A}} = \frac{1}{e_{\mbox{K}}} {\bf P}_{\mbox{C}} . 
\end{equation} 
The positional vector ${\bf P}_{\mbox{A}}$ is still located 
on the apparent ellipse given by Eq. $(\ref{ellipse3})$. 
We thus obtain the ellipticity as 
\begin{equation}
e_{\mbox{K}}=\sqrt{\frac{x_{\mbox{c}}^2}{a^2}+\frac{y_{\mbox{c}}^2}{b^2}} . 
\label{eK}
\end{equation}

\subsection{Projection onto a celestial sphere}
In the original derivation of AAK formula, the fact that 
the semimajor and semiminor axes divide the area of 
the ellipse in quarters. 
This still holds even after the projection. 
Namely the projected semimajor and semiminor axes divide the area of 
the apparent ellipse in quarters, though the original 
semimajor and semiminor axes are not always projected onto 
the apparent semimajor and semiminor ones. 
This way of the derivation, however, can be used for 
neither hyperbolic nor parabolic cases, where there are no counterparts 
of the semiminor axis. Hence we shall employ another method. 

In this paragraph, we use the Cartesian coordinates $(X, Y)$ 
on the original orbital plane. 
Let $i$ be the inclination angle between the original orbital
plane and the celestial sphere. We define as $\omega$ 
the angular distance of the periastron, namely the angle 
between the periastron and the ascending node. 
Let us express the original Keplerian ellipse as 
\begin{equation}
\frac{X^2}{a_{\mbox{K}}^2}+\frac{Y^2}{b_{\mbox{K}}^2}=1 .  
\label{originalellipse}
\end{equation}
We consider the line that is perpendicular to the semimajor axis 
at a focus $(a_{\mbox{K}}e_{\mbox{K}}, 0)$. 
This line intersects the original ellipse at points 
${\bf Q}=(a_{\mbox{K}}e_{\mbox{K}}, 
a_{\mbox{K}}(1-e_{\mbox{K}}^2))$
and 
${\bf R}=(a_{\mbox{K}}e_{\mbox{K}},
-a_{\mbox{K}}(1-e_{\mbox{K}}^2))$. 
For later convenience, we adopt the coordinates 
$(\bar{X}, \bar{Y})$ whose origin is located at the focus, 
by making a translation as $\bar{X}=X-a_{\mbox{K}}e_{\mbox{K}}$. 
Then, one rewrites  
${\bf Q}=(0, 
a_{\mbox{K}}(1-e_{\mbox{K}}^2))$
and 
${\bf R}=(0,
-a_{\mbox{K}}(1-e_{\mbox{K}}^2))$.

Only in this paragraph, we adopt other Cartesian coordinates 
$(x', y')$ so that the ascending node can be located on the $x'$-axis 
and the origin can be the common center of mass.  
The true periastron of the original ellipse is projected at 
${\bf P}_{\mbox{A}}\equiv(x_{\mbox{A}}', y_{\mbox{A}}')
=a_{\mbox{K}}(1-e_{\mbox{K}})(\cos\omega, \sin\omega\cos i)$. 
The point denoted by ${\bf Q}$ is projected at 
${\bf P}_{\mbox{Q}}\equiv(x_{\mbox{Q}}', y_{\mbox{Q}}')
=a_{\mbox{K}}(1-e_{\mbox{K}}^2) (-\sin\omega, \cos\omega\cos i)$.
It is useful to consider the following invariants because 
the components of a vector depend on the adopted coordinates. 
\begin{eqnarray}
&&|{\bf P}_{\mbox{A}}|^2=a_{\mbox{K}}^2(1-e_{\mbox{K}})^2
(\cos^2\omega+\sin^2\omega\cos^2 i) , 
\label{PA}\\
&&|{\bf P}_{\mbox{Q}}|^2=a_{\mbox{K}}^2(1-e_{\mbox{K}}^2)^2
(\sin^2\omega+\cos^2\omega\cos^2 i) , 
\label{PQ}\\
&&|{\bf P}_{\mbox{A}}\times{\bf P}_{\mbox{Q}}|
=a_{\mbox{K}}^2(1+e_{\mbox{K}})(1-e_{\mbox{K}})^2\cos i . 
\label{times}
\end{eqnarray}

We consider the area surrounded by the ellipse and the line interval 
between ${\bf Q}$ and ${\bf R}$. This area is divided into equal
halves by the semimajor axis. Even after the projection, the 
divided areas are still equal halves. Hence one can determine 
the location of the projected ${\bf Q}$ as  
\begin{equation}
{\bf P}_{\mbox{Q}} = 
\left(x_{\mbox{c}}-\frac{ay_{\mbox{c}}s}{b}, 
y_{\mbox{c}}+\frac{bx_{\mbox{c}}s}{a}\right) , 
\end{equation}
in the apparent ellipse coordinates, where we defined 
\begin{equation}
s=\frac{\sqrt{1-e_{\mbox{K}}^2}}{e_{\mbox{K}}} .
\end{equation}
In this computation, it is useful to stretch the apparent 
ellipse along its semiminor axis by $a/b$ so that one can 
consider a circle with radius $a$. 
In this stretching, importantly, the areal division 
into equal halves still holds. 
We make a translation as $x \to x-x_{\mbox{c}}$ and 
$y \to y-y_{\mbox{c}}$ so that the projected common center 
of mass can become the origin of the new coordinates.  
Then, we have 
\begin{equation}
{\bf P}_{\mbox{A}} = \frac{1-e_{\mbox{K}}}{e_{\mbox{K}}} 
(x_{\mbox{c}}, y_{\mbox{c}}) , 
\end{equation}
\begin{equation}
{\bf P}_{\mbox{Q}} = 
\left(-\frac{ay_{\mbox{c}}s}{b}, \frac{bx_{\mbox{c}}s}{a}\right) . 
\end{equation}
Hence, we obtain the invariants from these vectors as 
\begin{eqnarray}
&&|{\bf P}_{\mbox{A}}|^2=
\left(\frac{1-e_{\mbox{K}}}{e_{\mbox{K}}}\right)^2 
(x_{\mbox{c}}^2 + y_{\mbox{c}}^2) , 
\label{PA2}\\
&&|{\bf P}_{\mbox{Q}}|^2=
\frac{(a^4y_{\mbox{c}}^2 + b^4x_{\mbox{c}}^2)(1-e_{\mbox{K}}^2)}
{a^2b^2e_{\mbox{K}}^2} , 
\label{PQ2}\\
&&|{\bf P}_{\mbox{A}}\times{\bf P}_{\mbox{Q}}|
=ab (1-e_{\mbox{K}}) \sqrt{1-e_{\mbox{K}}^2} , 
\label{times2}
\end{eqnarray}
whose values can be estimated because $a$, $b$, $x_{\mbox{c}}$, 
$y_{\mbox{c}}$ and $e_{\mbox{K}}$ have been already all determined 
up to this point. 

Equations ($\ref{PA}$)-($\ref{times}$) for $\cos i$, 
$a_{\mbox{K}}$ and $\cos 2\omega$ are solved as 
\begin{eqnarray}
\cos i&=&\frac12 (\xi - \sqrt{\xi^2-4}) ,
\label{cosi}\\
a_{\mbox{K}}&=&\frac{1}{1-e_{\mbox{K}}^2}
\sqrt{\frac{(1+e_{\mbox{K}})^2 |{\bf P}_{\mbox{A}}|^2 
+ |{\bf P}_{\mbox{Q}}|^2}{1+\cos^2 i}} , 
\label{aK}\\
\cos 2\omega&=&\frac{(1+e_{\mbox{K}})^2 |{\bf P}_{\mbox{A}}|^2 
- |{\bf P}_{\mbox{Q}}|^2} 
{a_{\mbox{K}}^2 (1-e_{\mbox{K}}^2)^2 \sin^2 i} , 
\label{cos2omega}
\end{eqnarray}
where we define 
\begin{equation}
\xi=\frac{(1+e_{\mbox{K}})^2 |{\bf P}_{\mbox{A}}|^2 
+ |{\bf P}_{\mbox{Q}}|^2}
{(1+e_{\mbox{K}}) |{\bf P}_{\mbox{A}}\times{\bf Q}_{\mbox{Q}}|} . 
\label{xi}
\end{equation}
One can show 
\begin{equation}
\xi \geq 2 ,  
\label{xi2} 
\end{equation}
because the arithmetic mean is not smaller than the geometric one. 
It is worthwhile to mention that Eq. ($\ref{cosi}$) is 
obtained by solving a quadratic equation for $\cos i$ as 
\begin{equation}
\cos^2 i-\xi \cos i+1=0 , 
\label{cos2}
\end{equation}
which can be obtained from Eqs. ($\ref{PA}$)-($\ref{times}$) 
by eliminating $\omega$ and $a_{\mbox{K}}$. 
Furthermore, one can prove that a root of 
$\cos i=(\xi + \sqrt{\xi^2-4})/2$ 
must be abandoned because Eq. ($\ref{xi2}$) implies that 
it is always larger than the unity. 
Only in the case of $i=0$, the apparent ellipse coincides with 
the true orbit. 
Hence, the ascending node and consequently the angular distance of 
the periastron make no sense. 
As a result, the denominator of R. H. S. of Eq. ($\ref{cos2omega}$) 
vanishes. 

Equations ($\ref{eK}$), ($\ref{cosi}$), ($\ref{aK}$) 
and ($\ref{cos2omega}$) agree with those of AAK, where 
different notations were employed. 
In this paper, the semiminor axis is not used for areal divisions. 
Therefore, this formalism can be generalized straightforwardly 
to an open orbit, as shown below.

\section{The solution for a hyperbolic orbit}
\subsection{An apparent hyperbola}
Let a star move in a hyperbola on a celestial sphere. 
Without loss of generality, we can assume that 
the hyperbola is expressed as 
\begin{equation}
\frac{x^2}{a^2}-\frac{y^2}{b^2}=1 , 
\label{hyperbola}
\end{equation}
and the orbit is the left-hand side of the hyperbola, $x<0$. 
Then the position of the star at each epoch $t_j$ is denoted by 
\begin{eqnarray} 
{\bf P}_j&=&(x_j, y_j) 
\nonumber\\ 
&=&(-a \cosh u_j, b \sinh u_j) . 
\label{star-h}
\end{eqnarray}
The projected common center of mass 
${\bf P}_{\mbox{C}}\equiv (x_{\mbox{C}}, y_{\mbox{C}})$ is 
not necessarily a focus of the apparent hyperbola but 
the projected focus of the original Keplerian hyperbola. 

The projected areal velocity with respect to the projected 
common center of mass is denoted by $dS/dt$. 
The law of the constant areal velocity on the observed plane 
is written as 
\begin{equation}
\frac{dS}{dt}=\frac{S(j, k)}{T(j, k)} , 
\label{areavelocity-h}
\end{equation}
where for $t_j > t_k$ we obtain 
\begin{eqnarray}
S(j, k)=-\frac12 ab  
\Bigl[
u_j-u_k
+\frac{x_{\mbox{c}}}{a}(\sinh u_j-\sinh u_k)
+\frac{y_{\mbox{c}}}{b}(\cosh u_j-\cosh u_k) 
\Bigr] . 
\label{areaS-h}
\end{eqnarray}

Equation $(\ref{areavelocity-h})$ is rewritten explicitly as 
\begin{eqnarray}
\frac{S(2, 1)}{T(2, 1)}&=&\frac{S(3, 2)}{T(3, 2)} , \\
\frac{S(3, 2)}{T(3, 2)}&=&\frac{S(4, 3)}{T(4, 3)} . 
\label{areavelocity1234-h}
\end{eqnarray}
They are solved for $x_{\mbox{c}}$ and $y_{\mbox{c}}$ as 
\begin{eqnarray}
x_{\mbox{c}}&=&a \frac{E_1 F_2-E_2 F_1}{D_1 E_2-D_2 E_1} , 
\label{xC-h}\\
y_{\mbox{c}}&=&b \frac{F_1 D_2-F_2 D_1}{D_1 E_2-D_2 E_1} , 
\label{yC-h}
\end{eqnarray}
where 
\begin{eqnarray}
D_j&=&T(j+1, j)\sinh u_{j+2}+T(j+2, j+1)\sinh u_j
\nonumber\\
&&
-T(j+2, j)\sinh u_{j+1} , 
\label{Dj} \\
E_j&=&T(j+1, j)\cosh u_{j+2}+T(j+2, j+1)\cosh u_j
\nonumber\\
&&
-T(j+2, j)\cosh u_{j+1} , 
\label{Ej} \\
F_j&=&T(j+1, j) u_{j+2}+T(j+2, j+1) u_j-T(j+2, j) u_{j+1} . 
\label{Fj} 
\end{eqnarray}

The periastron is projected onto the observed hyperbola 
at ${\bf P}_{\mbox{A}}\equiv(x_{\mbox{A}}, y_{\mbox{A}})$. 
The ratio of the semimajor axis to the distance between the center  
and the focus of the hyperbola remains the same, 
even after the projection. 
Hence, we find 
\begin{equation}
{\bf P}_{\mbox{A}} = \frac{1}{e_{\mbox{K}}} {\bf P}_{\mbox{C}} . 
\label{PA-h2}
\end{equation} 
The positional vector ${\bf P}_{\mbox{A}}$ is still located 
on the apparent hyperbola given by Eq. $(\ref{hyperbola})$. 
We thus obtain the ellipticity as 
\begin{equation}
e_{\mbox{K}}=\sqrt{\frac{x_{\mbox{c}}^2}{a^2}-\frac{y_{\mbox{c}}^2}{b^2}} . 
\label{eK-h}
\end{equation}

\subsection{Projection onto a celestial sphere}
In this paragraph, we use the Cartesian coordinates $(X, Y)$ 
on the original orbital plane. 
Let us express an original Keplerian hyperbola as 
\begin{equation}
\frac{X^2}{a_{\mbox{K}}^2}-\frac{Y^2}{b_{\mbox{K}}^2}=1 .  
\label{originalhyperbola}
\end{equation}
We consider the line that is perpendicular to the semimajor axis 
at a focus $(a_{\mbox{K}}e_{\mbox{K}}, 0)$. 
This line intersects the original hyperbola at 
${\bf Q}=(a_{\mbox{K}}e_{\mbox{K}}, 
a_{\mbox{K}}(e_{\mbox{K}}^2-1))$
and 
${\bf R}=(a_{\mbox{K}}e_{\mbox{K}},
-a_{\mbox{K}}(e_{\mbox{K}}^2-1))$.

Only in this paragraph, we shall employ other Cartesian coordinates 
$(x', y')$ so that the ascending node can be located on the $x'$-axis 
and the origin can be the common center of mass.  
The true periastron of the original hyperbola is projected at 
${\bf P}_{\mbox{A}}\equiv(x_{\mbox{A}}', y_{\mbox{A}}')
=a_{\mbox{K}}(e_{\mbox{K}}-1)(\cos\omega, \sin\omega\cos i)$, 
where $i$ and $\omega$ are the inclination angle and 
the angular distance of the periastron, respectively. 
The point denoted by ${\bf Q}$ is projected at 
${\bf P}_{\mbox{Q}}\equiv(x_{\mbox{Q}}', y_{\mbox{Q}}')
=a_{\mbox{K}}(e_{\mbox{K}}^2-1) (-\sin\omega, \cos\omega\cos i)$.
We shall use the following invariants as 
\begin{eqnarray}
&&|{\bf P}_{\mbox{A}}|^2=a_{\mbox{K}}^2(e_{\mbox{K}}-1)^2
(\cos^2\omega+\sin^2\omega\cos^2 i) , 
\label{PA-h}\\
&&|{\bf P}_{\mbox{Q}}|^2=a_{\mbox{K}}^2(e_{\mbox{K}}^2-1)^2
(\sin^2\omega+\cos^2\omega\cos^2 i) , 
\label{PQ-h}\\
&&|{\bf P}_{\mbox{A}}\times{\bf P}_{\mbox{Q}}|
=a_{\mbox{K}}^2(e_{\mbox{K}}+1)(e_{\mbox{K}}-1)^2\cos i . 
\label{times-h}
\end{eqnarray}

We consider the area surrounded by the hyperbola and the line interval 
between ${\bf Q}$ and ${\bf R}$. This area is divided into equal
halves by the semimajor axis. Even after the projection, 
the divided areas are still equal halves.
Hence one can determine the location of the projected ${\bf Q}$ as  
\begin{equation}
{\bf P}_Q = 
\left(x_{\mbox{c}}-\frac{ay_{\mbox{c}}s_{\mbox{h}}}{b}, 
y_{\mbox{c}}+\frac{bx_{\mbox{c}}s_{\mbox{h}}}{a}\right) , 
\label{PQ-h2}
\end{equation}
in the apparent hyperbola coordinates, where we defined 
\begin{equation}
s_{\mbox{h}}=\frac{\sqrt{e_{\mbox{K}}^2-1}}{e_{\mbox{K}}} .
\end{equation}
We make a translation as $x \to x-x_{\mbox{c}}$ and 
$y \to y-y_{\mbox{c}}$ so that the center of the coordinates 
can move to the projected common center of mass. 
Then, we have
\begin{equation}
{\bf P}_{\mbox{A}} = \frac{1-e_{\mbox{K}}}{e_{\mbox{K}}} 
(x_{\mbox{c}}, y_{\mbox{c}}) , 
\end{equation}
\begin{equation}
{\bf P}_Q = 
\left(-\frac{ay_{\mbox{c}}s_{\mbox{h}}}{b}, 
\frac{bx_{\mbox{c}}s_{\mbox{h}}}{a}\right) , 
\end{equation}
where we used Eqs. ($\ref{PA-h2}$) and ($\ref{PQ-h2}$). 
Hence, we obtain the invariants from these vectors as 
\begin{eqnarray}
&&|{\bf P}_{\mbox{A}}|^2=
\left(\frac{e_{\mbox{K}}-1}{e_{\mbox{K}}}\right)^2 
(x_{\mbox{c}}^2 + y_{\mbox{c}}^2) , 
\label{PA2-h}\\
&&|{\bf P}_{\mbox{Q}}|^2=
\frac{(a^4y_{\mbox{c}}^2 + b^4x_{\mbox{c}}^2)(e_{\mbox{K}}^2-1)}
{a^2b^2e_{\mbox{K}}^2} , 
\label{PQ2-h}\\
&&|{\bf P}_{\mbox{A}}\times{\bf P}_{\mbox{Q}}|
=ab (e_{\mbox{K}}-1) \sqrt{e_{\mbox{K}}^2-1} , 
\label{times2-h}
\end{eqnarray}
whose values can be estimated because $a$, $b$, $x_{\mbox{c}}$, 
$y_{\mbox{c}}$ and $e_{\mbox{K}}$ have been all determined 
up to this point.

Equations ($\ref{PA-h}$)-($\ref{times-h}$) for $\cos i$, 
$a_{\mbox{K}}$ and $\cos 2\omega$ are solved as 
\begin{eqnarray}
\cos i&=&\frac12 \left(\xi - \sqrt{\xi^2-4}\right) ,
\label{cosi-h}\\
a_{\mbox{K}}&=&\frac{1}{e_{\mbox{K}}^2-1}
\sqrt{\frac{(e_{\mbox{K}}+1)^2 |{\bf P}_{\mbox{A}}|^2 
+ |{\bf P}_{\mbox{Q}}|^2}{1+\cos^2 i}} , 
\label{aK-h}\\
\cos 2\omega&=&\frac{(e_{\mbox{K}}+1)^2 |{\bf P}_{\mbox{A}}|^2 
- |{\bf P}_{\mbox{Q}}|^2} 
{a_{\mbox{K}}^2 (e_{\mbox{K}}^2-1)^2 \sin^2 i} , 
\label{cos2omega-h}
\end{eqnarray}
where we define
\begin{equation}
\xi=\frac{(e_{\mbox{K}}+1)^2 |{\bf P}_{\mbox{A}}|^2 
+ |{\bf P}_{\mbox{Q}}|^2}
{(e_{\mbox{K}}+1) |{\bf P}_{\mbox{A}}\times{\bf Q}_{\mbox{Q}}|} . 
\label{xi-h}
\end{equation}
In the similar manner to the elliptic case, one can show 
\begin{equation}
\xi \geq 2 . 
\label{xi2-h} 
\end{equation}
Hence, for a quadratic equation for $\cos i$ as 
\begin{equation}
\cos^2 i-\xi \cos i+1=0 , 
\label{cos2-h}
\end{equation}
which is derived from Eqs. ($\ref{PA-h}$)-($\ref{times-h}$).  
One can prove that a root of $\cos i=(\xi + \sqrt{\xi^2-4})/2$ 
is larger than the unity according to Eq. ($\ref{xi2-h}$) 
and thus must be abandoned.

\section{The solution for a parabolic orbit}
\subsection{An apparent parabola}
Let a star move in a parabola 
on a celestial sphere as 
\begin{equation}
y^2 + 4qx=0 . 
\label{parabola}
\end{equation}
Then the position of the star at each epoch $t_j$ is 
denoted by 
\begin{eqnarray} 
{\bf P}_j&=&(x_j, y_j) 
\nonumber\\ 
&=&(-\frac12 q u_j^2, \sqrt{2} q u_j) . 
\label{star-p}
\end{eqnarray}
The projected common center of mass 
${\bf P}_{\mbox{C}}\equiv (x_{\mbox{C}}, y_{\mbox{C}})$ is 
not necessarily a focus of the apparent parabola but 
the projected focus of the original Keplerian parabola. 

The law of the constant areal velocity on the observed plane 
is written as 
\begin{equation}
\frac{dS}{dt}=\frac{S(j, k)}{T(j, k)} , 
\label{areavelocity-p}
\end{equation}
where for $t_j > t_k$ we obtain 
\begin{eqnarray}
S(j, k)&=&-\frac13 (\sqrt{-x_j}-\sqrt{-x_k}) 
\Bigl[
\sqrt{q}
(x_j-\sqrt{x_jx_k}+x_k) 
\nonumber\\
&&+\frac32 (\sqrt{-x_j}+\sqrt{-x_k}) y_{\mbox{C}} 
+3 \sqrt{q} x_{\mbox{C}} 
\Bigr] . 
\label{areaS-p}
\end{eqnarray}

Equation $(\ref{areavelocity-p})$ is rewritten explicitly as 
\begin{eqnarray}
\frac{S(2, 1)}{T(2, 1)}&=&\frac{S(3, 2)}{T(3, 2)} , \\
\frac{S(3, 2)}{T(3, 2)}&=&\frac{S(4, 3)}{T(4, 3)} . 
\label{areavelocity1234-p}
\end{eqnarray}
They are solved for $x_{\mbox{c}}$ and $y_{\mbox{c}}$ as 
\begin{eqnarray}
x_{\mbox{c}}&=&- \frac{H_1 I_2-H_2 I_1}{G_1 H_2-G_2 H_1} , 
\label{xC-p}\\
y_{\mbox{c}}&=&-\sqrt{q} \frac{G_1 I_2-G_2 I_1}{G_1 H_2-G_2 H_1} , 
\label{yC-p}
\end{eqnarray}
where 
\begin{eqnarray}
G_j&=&3 [T(j+1, j)\sqrt{-x_{j+2}}+T(j+2, j+1)\sqrt{-x_j} 
\nonumber\\
&&-T(j+2, j)\sqrt{-x_{j+1}}] , 
\label{Gj} \\
H_j&=&\frac32 [T(j+1, j)x_{j+2}+T(j+2, j+1)x_j 
-T(j+2, j)x_{j+1}] , 
\label{Hj} \\
I_j&=&- [T(j+1, j)x_{j+2}\sqrt{-x_{j+2}}+T(j+2, j+1)x_j\sqrt{-x_j} 
\nonumber\\
&&-T(j+2, j)x_{j+1}\sqrt{-x_{j+1}}] . 
\label{Ij} 
\end{eqnarray}

The periastron is projected onto the observed parabola 
at ${\bf P}_{\mbox{A}}\equiv(x_{\mbox{A}}, y_{\mbox{A}})$. 
The semimajor axis is projected onto a line, which may be  
expressed as $y=K x+ L$. 
This line intersects the apparent parabola only at 
the projected periastron. 
Therefore, we find $K=0$ because $|x|$ must be larger than 
$\sqrt{-x}$ for a sufficient large $|x|$. 
In addition, the projected semimajor axis goes through 
the projected common center. 
This implies $L=y_{\mbox{c}}$. 
Hence we obtain 
\begin{equation}
{\bf P}_{\mbox{A}}
=\left(-\frac{y_{\mbox{c}}^2}{4q}, y_{\mbox{c}}\right). 
\end{equation}

\subsection{Projection onto a celestial sphere}
In this paragraph, we use the Cartesian coordinates $(X, Y)$ 
on the original orbital plane. 
Let a Keplerian parabola be  
\begin{equation}
Y^2+4 q_{\mbox{K}} X=0 .   
\label{originalparabola}
\end{equation}
We consider the line that is perpendicular to the semimajor axis 
at a focus $(-q_{\mbox{K}}, 0)$. 
This line intersects the original parabola at 
${\bf Q}=(-q_{\mbox{K}}, 2 q_{\mbox{K}})$
and 
${\bf R}=(-q_{\mbox{K}}, -2 q_{\mbox{K}})$.

Only in this paragraph, we employ other Cartesian coordinates 
$(x', y')$ so that the ascending node can be located on the $x'$-axis 
and the origin can be the common center of mass.  
The true periastron of the original parabola is projected at 
${\bf P}_{\mbox{A}}\equiv(x_{\mbox{A}}', y_{\mbox{A}}')
=q_{\mbox{K}} (\cos\omega, \sin\omega\cos i)$, 
where $i$ and $\omega$ are the inclination angle and 
the angular distance of the periastron, respectively. 
The point denoted by ${\bf Q}$ is projected at 
${\bf P}_{\mbox{Q}}\equiv(x_{\mbox{Q}}', y_{\mbox{Q}}')
=2 q_{\mbox{K}} (-\sin\omega, \cos\omega\cos i)$.
We shall use the following invariants as 
\begin{eqnarray}
&&|{\bf P}_{\mbox{A}}|^2=q_{\mbox{K}}^2
(\cos^2\omega+\sin^2\omega\cos^2 i) , 
\label{PA-p}\\
&&|{\bf P}_{\mbox{Q}}|^2=4 q_{\mbox{K}}^2
(\sin^2\omega+\cos^2\omega\cos^2 i) , 
\label{PQ-p}\\
&&|{\bf P}_{\mbox{A}}\times{\bf P}_{\mbox{Q}}|
=2 q_{\mbox{K}}^2 \cos i . 
\label{times-p}
\end{eqnarray}

We consider the area surrounded by the parabola and the line interval 
between ${\bf Q}$ and ${\bf R}$. This area is divided into equal
halves by the semimajor axis. Even after the projection, 
the divided areas are still equal halves. 
Hence one can determine the location of the projected ${\bf Q}$ as  
\begin{equation}
{\bf P}_Q = 
\left(x_{\mbox{c}}-\frac{y_{\mbox{c}}s_{\mbox{p}}}{2q}, 
y_{\mbox{c}}+s_{\mbox{p}}\right) , 
\end{equation}
in the apparent parabola coordinates, where we defined 
\begin{equation}
s_{\mbox{p}}=\sqrt{-(y_{\mbox{c}}^2+4q x_{\mbox{c}})} .
\end{equation}
We make a translation as $x \to x-x_{\mbox{c}}$ and 
$y \to y-y_{\mbox{c}}$ so that the origin of the coordinates 
can be the projected common center of mass. 
Then, we obtain 
\begin{equation}
{\bf P}_{\mbox{A}}
=\left(-\frac{y_{\mbox{c}}^2+4qx_{\mbox{c}}}{4q}, 0\right) , 
\end{equation}
\begin{equation}
{\bf P}_Q = 
\left(-\frac{y_{\mbox{c}}s_{\mbox{p}}}{2q}, 
s_{\mbox{p}}\right) . 
\end{equation}
Hence, we obtain the invariants from these vectors as 
\begin{eqnarray}
&&|{\bf P}_{\mbox{A}}|^2=
\frac{(y_{\mbox{c}}^2+4qx_{\mbox{c}})^2}{16q^2} , 
\label{PA2-p}\\
&&|{\bf P}_{\mbox{Q}}|^2=
-\frac{(y_{\mbox{c}}^2+4qx_{\mbox{c}}) (y_{\mbox{c}}^2+4q^2)}
{4q^2} , 
\label{PQ2-p}\\
&&|{\bf P}_{\mbox{A}}\times{\bf P}_{\mbox{Q}}|
=\frac{[-(y_{\mbox{c}}^2+4qx_{\mbox{c}})]^{3/2}}{4q} , 
\label{times2-p}
\end{eqnarray}
whose values can be estimated because $q$, $x_{\mbox{c}}$ and  
$y_{\mbox{c}}$ have been all determined 
up to this point.

Equations ($\ref{PA-p}$)-($\ref{times-p}$) for $\cos i$, 
$a_{\mbox{K}}$ and $\cos 2\omega$ are solved as 
\begin{eqnarray}
\cos i&=&\frac12 (\xi - \sqrt{\xi^2-4}) ,
\label{cosi-p}\\
q_{\mbox{K}}&=&\frac12
\sqrt{\frac{4 |{\bf P}_{\mbox{A}}|^2 
+ |{\bf P}_{\mbox{Q}}|^2}{1+\cos^2 i}} , 
\label{aK-p}\\
\cos 2\omega&=&\frac{4 |{\bf P}_{\mbox{A}}|^2 
- |{\bf P}_{\mbox{Q}}|^2} 
{4 q_{\mbox{K}}^2 \sin^2 i} , 
\label{cos2omega-p}
\end{eqnarray}
where we define
\begin{equation}
\xi=\frac{4 |{\bf P}_{\mbox{A}}|^2 
+ |{\bf P}_{\mbox{Q}}|^2}
{2 |{\bf P}_{\mbox{A}}\times{\bf Q}_{\mbox{Q}}|} . 
\label{xi-p}
\end{equation}
In the similar manner to the above two cases, one can show 
\begin{equation}
\xi \geq 2 . 
\label{xi2-p} 
\end{equation}
Hence, for a quadratic equation for $\cos i$ as 
\begin{equation}
\cos^2 i-\xi \cos i+1=0 , 
\label{cos2-p}
\end{equation}
which is derived from Eqs. ($\ref{PA-p}$)-($\ref{times-p}$). 
One can prove that a root of $\cos i=(\xi + \sqrt{\xi^2-4})/2$ 
is larger than the unity according to Eq. ($\ref{xi2-p}$) 
and thus must be abandoned.

\section{Transformations from an elliptic case to 
hyperbolic/parabolic ones} 
\subsection{To a hyperbolic case} 
Let us rederive the formula for a hyperbolic case from that 
for an elliptic one by making a transformation as 
\begin{eqnarray}
u&\to&\hat i u ,
\label{tr-u-h}
\\
a&\to&-a ,
\label{tr-a-h}
\\
b&\to&-\hat i b ,
\label{tr-b-h}
\end{eqnarray}
which imply 
\begin{eqnarray}
\cos u&\to&\cosh u ,
\label{tr-cosu-h}
\\
\sin u&\to&\hat i\sinh u ,
\label{tr-sinu-h}
\end{eqnarray}
where $\hat i=\sqrt{-1}$. 
Then, from Eqs. ($\ref{Aj}$)-($\ref{Cj}$) and 
($\ref{Dj}$)-($\ref{Fj}$) we find 
\begin{eqnarray}
A_j&\to&\hat i D_j , 
\\
B_j&\to& E_j , 
\\
C_j&\to&\hat i F_j .
\end{eqnarray}
We can thus show that the location of the common center 
is transformed from Eqs. ($\ref{xC}$) and ($\ref{yC}$) 
to Eqs. ($\ref{xC-h}$) and ($\ref{yC-h}$). 
Equation ($\ref{eK}$) is transformed into Eq. ($\ref{eK-h}$),  
Eqs. ($\ref{PA2}$)-($\ref{times2}$) into 
Eqs. ($\ref{PA2-h}$)-($\ref{times2-h}$), 
Eqs. ($\ref{cosi}$)-($\ref{cos2omega}$) 
into Eqs. ($\ref{cosi-h}$)-($\ref{cos2omega-h}$), 
because $\xi$ remains unchanged.

\subsection{To a parabolic case} 
To rederive the formula for a parabolic case, we perform 
a transformation from an elliptic case with a limiting procedure as  
\begin{eqnarray}
e&\to&1 ,
\label{tr-e-p}
\\
q&=&\lim_{e\to 1} a(1-e) ,
\label{tr-q-p}
\\
x^{\prime}&=&x-a ,
\label{tr-x-p}
\end{eqnarray}
where the finite $q$ implies $a \to \infty$ and 
$b^2=a(1+e)\times a(1-e) \to 2aq$. 
Then, we find 
\begin{eqnarray}
A_j&\to&\frac{2\sqrt{q}}{3b} G_j , 
\\
B_j&\to&\frac{2}{3a} H_j , 
\\
C_j&\to&\frac{2\sqrt{q}}{3b} G_j+\frac{4q^{3/2}}{3b^3} I_j .
\end{eqnarray}
We can transform the location of the common center from 
Eqs. ($\ref{xC}$) and ($\ref{yC}$) to  
\begin{eqnarray}
x_{\mbox{c}}&\to&a-\frac{H_1 I_2-H_2 I_1}{G_1 H_2-G_2 H_1} , 
\label{tr-xC-p}\\
y_{\mbox{c}}&\to&
-\sqrt{q} \frac{G_1 I_2-G_2 I_1}{G_1 H_2-G_2 H_1} , 
\label{tr-yC-p}
\end{eqnarray}
which agrees with Eq. ($\ref{yC-p}$). 
We thus recover Eq. ($\ref{xC-p}$) as
\begin{eqnarray}
x^{\prime}_{\mbox{c}}
&\to&-\frac{H_1 I_2-H_2 I_1}{G_1 H_2-G_2 H_1} . 
\label{tr-xC2-p}
\end{eqnarray}

Equation ($\ref{eK}$) is transformed as 
\begin{eqnarray}
e_{\mbox{K}}&=&
1+\frac{y_{\mbox{c}}^2+4qx^{\prime}_{\mbox{c}}}{4aq}
+O\left(\frac{1}{a^2}\right)
\nonumber\\
&\to&1 , 
\label{tr-eK-p}
\end{eqnarray}
where we used $a \to \infty$ and $b^2 \to 2aq$.
By using Eq. ($\ref{tr-eK-p}$) and $b^2 \to 2aq$, we obtain 
\begin{eqnarray}
|{\bf P}_{\mbox{A}}|^2
&\to&
\frac{(y_{\mbox{c}}^2+4qx^{\prime}_{\mbox{c}})^2 }{16q^2} , 
\label{tr-PA2-p}\\
|{\bf P}_{\mbox{Q}}|^2&\to&
\frac{2[a^4y_{\mbox{c}}^2 + 4a^2q^2 (a+x^{\prime}_{\mbox{c}})^2]
(1-e_{\mbox{K}})}
{2a^3q}
\nonumber\\
&\to&
-\frac{(y_{\mbox{c}}^2 + 4q^2)
(y_{\mbox{c}}^2+4qx^{\prime}_{\mbox{c}})}
{4q^2} , 
\label{tr-PQ2-p}\\
|{\bf P}_{\mbox{A}}\times{\bf P}_{\mbox{Q}}|
&=&2 a^{3/2}\sqrt{q} (1-e_{\mbox{K}})^{3/2}
\nonumber\\
&\to&\frac{[-(y_{\mbox{c}}^2+4qx^{\prime}_{\mbox{c}})]^{3/2}}{4q} , 
\label{tr-times2-p}
\end{eqnarray}
which agree with Eqs. ($\ref{PA2-p}$)-($\ref{times2-p}$). 

Equations ($\ref{cosi}$)-($\ref{cos2omega}$) are transformed as 
\begin{eqnarray}
\cos i&=&\frac12 (\xi - \sqrt{\xi^2-4}) , 
\label{tr-cosi-p}\\
q_{\mbox{K}}&=&
\lim_{e_{\mbox{K}}\to 1} a_{\mbox{K}} (1-e_{\mbox{K}}) 
\nonumber\\
&=&\lim_{e_{\mbox{K}}\to 1} 
\frac{1}{1+e_{\mbox{K}}}
\sqrt{\frac{(1+e_{\mbox{K}})^2 |{\bf P}_{\mbox{A}}|^2 
+ |{\bf P}_{\mbox{Q}}|^2}{1+\cos^2 i}} 
\nonumber\\
&=&
\frac12
\sqrt{\frac{4 |{\bf P}_{\mbox{A}}|^2 
+ |{\bf P}_{\mbox{Q}}|^2}{1+\cos^2 i}} , 
\label{tr-aK-p}\\
\cos 2\omega&=&\frac{(1+e_{\mbox{K}})^2 |{\bf P}_{\mbox{A}}|^2 
- |{\bf P}_{\mbox{Q}}|^2} 
{a_{\mbox{K}}^2 (1-e_{\mbox{K}})^2 (1+e_{\mbox{K}})^2 \sin^2 i} 
\nonumber\\
&\to&\frac{4 |{\bf P}_{\mbox{A}}|^2 
- |{\bf P}_{\mbox{Q}}|^2} 
{4 q_{\mbox{K}}^2 \sin^2 i} , 
\label{tr-cos2omega-p}
\end{eqnarray}
where $\xi$ remains unchanged. 
They agree with Eqs. ($\ref{cosi-p}$)-($\ref{cos2omega-p}$).

\section{Conclusion}
The formulae for orbit determination of elliptic, hyperbolic 
and parabolic orbits are obtained in a unified manner by generalizing 
AAK approach, which originally needed a fact of the areal divisions 
by the semimajor and semiminor axes of an ellipse. 
We show also that the present formulae are recovered from 
AAK result by a suitable transformation among an ellipse, 
hyperbola and parabola.

\section*{ACKNOWLEDGMENTS}
The present author would like to thank the anonymous reviewers 
for invaluable information particularly regarding the earlier works. 
He would like to thank Professor M. Kasai and Professor K. Maeda 
for encouragement.



%



\end{document}